\documentclass[a4paper,11pt]{article}
\pdfoutput=1 % if your are submitting a pdflatex (i.e. if you have
\usepackage{jinstpub} % for details on the use of the package, please
\usepackage{lineno}

\title{\boldmath Direct current measurements of the SPIDER beam: a comparison to existing beam diagnostics}

\author[a,b,1]{A. Shepherd,\note{Corresponding author.}}
\author[b]{T. Patton,}
\author[b]{A. Pimazzoni,}
\author[c,b]{B. Pouradier-Duteil,}
\author[b]{A. Rigoni Garola,}
\author[d,b]{E. Sartori,}
\author[e,b]{M. Ugoletti}
\author[b,f]{and G. Serianni}

\affiliation[a]{United Kingdom Atomic Energy Authority, Culham Centre for Fusion Energy,\\Culham Science Centre, Abingdon, Oxfordshire, OX14 3DB, UK}
\affiliation[b]{Consorzio RFX (CNR, ENEA, INFN, UNIPD, Acciaierie Venete SpA),\\Corso Stati Uniti 4 – 35127 Padova, Italy}
\affiliation[c]{Ecole Polytechnique Fédérale de Lausanne (EPFL),\\Swiss Plasma Center (SPC), 1015 Lausanne, Switzerland}
\affiliation[d]{Department of Management and Engineering, Università degli Studi di Padova,\\Strad. S. Nicola 3, 36100 Vicenza, Italy}
\affiliation[e]{INFN-LNL, Viale dell’Università 2, 35020 Legnaro, Italy}
\affiliation[f]{Institute for Plasma Science and Technologies of Consiglio Nazionale delle Ricerche, Corso Stati Uniti 4, 35127 Padova, Italy}

\emailAdd{alastair.shepherd@ukaea.uk}

\abstract{For negative ion beam sources there are several methods of measuring the accelerated beam current, most commonly electrical measurements at the power supply and calorimetric measurements. On SPIDER, the ITER Heating Neutral Beam full-scale beam source prototype, electrical measurements at the acceleration grid power supply (AGPS) are complemented by polarizing the diagnostic calorimeter STRIKE to provide an additional electrical measurement of the accelerated current. This is in addition to the calorimetric measurements provided by STRIKE. These diagnostics give differing measurements of the beam current. Exploiting the reduced number of open apertures on SPIDER a new beam diagnostic has been installed to measure the individual beamlet currents directly. The so called Beamlet Current Monitor (BCM) has been used to measure the current of five beamlets during the most recent SPIDER campaign. This work compares the BCM current to the electrical measurements at the AGPS and STRIKE. The average BCM current agrees well with the STRIKE electrical measurements, indicating that the AGPS overestimates the beam current. The individual beamlets are compared to the STRIKE calorimetric measurements, showing similar current trends with the source parameters.}


\proceeding{8$^{\text{th}}$ International symposium on Negative Ions, Beams and Sources - NIBS2022\\
  2-7 October 2022\\
  Orto Botanico - Padova, Italy}

\begin{document}
\maketitle
\flushbottom

\section{Introduction}
\label{sec:intro}

ITER is the next step on the roadmap to fusion energy, and requires a high energy Neutral Beam Injection (NBI) system to provide additional heating and current drive to the plasma. Two Heating Neutral Beams (HNBs) will be installed, providing 16.5 MW of heating power each, with a possible third HNB for a potential total of 50 MW \cite{Hemsworth_2017}. Negative ions of hydrogen or deuterium will be accelerated up to 1 MeV to achieve heating in the plasma core. The negative ions are produced in the source primarily through caesium-catalysed surface conversion of hydrogen/deuterium \cite{Bacal_2021}. To reach the ITER HNB requirements (1 MeV, 40 A in deuterium) the Neutral Beam Test Facility (NBTF) was established at Consorzio RFX, Italy \cite{Toigo_2017}.

The NBTF consists of two experiments: SPIDER, the full-scale ITER negative ion beam source prototype with 108 kV accelerator, and MITICA, the full-scale ITER HNB prototype with 1 MV accelerator. One of the aims of SPIDER is to achieve the ITER heating/diagnostic NBI requirements in extracted current density, uniformity and pulse length, (355 A/m$^{2}$ in hydrogen, $\pm10\%$, 3600 s) \cite{SERIANNI20192539}. SPIDER is equipped with an extensive array of diagnostics to measure the beam parameters \cite{Pasqualotto_2012,Pasqualotto_2017_2}. The extraction and acceleration power supplies provide measurements of the electron and ion currents. To observe the beam profile the Short-Time Retractable Instrumented Kalorimeter Experiment (STRIKE) is installed downstream of the accelerator, providing a measure of the spatial beam footprint as well as the accelerated current \cite{Pimazzoni_2020}. The beam width and intensity is also measured by beam tomography \cite{UGOLETTI2021112667} and Beam Emission Spectroscopy (BES) \cite{Barbisan_2021}.

Due to RF induced breakdowns on the back of the source, a mask has been installed downstream of the plasma grid (PG), to reduce the gas conductance through the accelerator and keep the vessel pressure below 40 mPa \cite{PAVEI2020112036}. This was sufficient to avoid the occurrence of breakdowns. Initially reducing the number of apertures from 1280 to 80, with a further reduction to 28 for the caesium campaign, the mask has allowed the study of individual beamlets. As well as an Allison emittance scanner (AES) \cite{Poggi_2020}, which measures the emittance and current of three beamlets, Beamlet Current Monitors (BCMs) have been installed downstream of the accelerator. Each BCM encircles a beamlet aperture and uses a combination of fluxgate (DC) and current transformer (AC) technologies to measure the beamlet current using the resultant magnetic flux. These sensors measure the current of five beamlets, from DC to several MHz AC \cite{Shepherd_2022,SHEPHERD2023113599,POURADIERDUTEIL2023113529}.

The BCMs were installed on SPIDER to provide a direct measurement of the accelerated beam current, as the existing beam current measurements are either electrical measurements that can include additional charged species (PS drain currents, STRIKE electrical) or are calculated indirectly (STRIKE IR calorimetry, tomography). This paper compares the different beam current measurements, in an attempt to characterise the differences and give a clear definition of the accelerated beam current. Where the diagnostics can resolve individual beamlets a comparison is made to validate the BCM measurements. 

\section{SPIDER beam diagnostics}
\label{sec:beam}

\begin{figure}[htpb]
\centering
\includegraphics[width=0.7\textwidth]{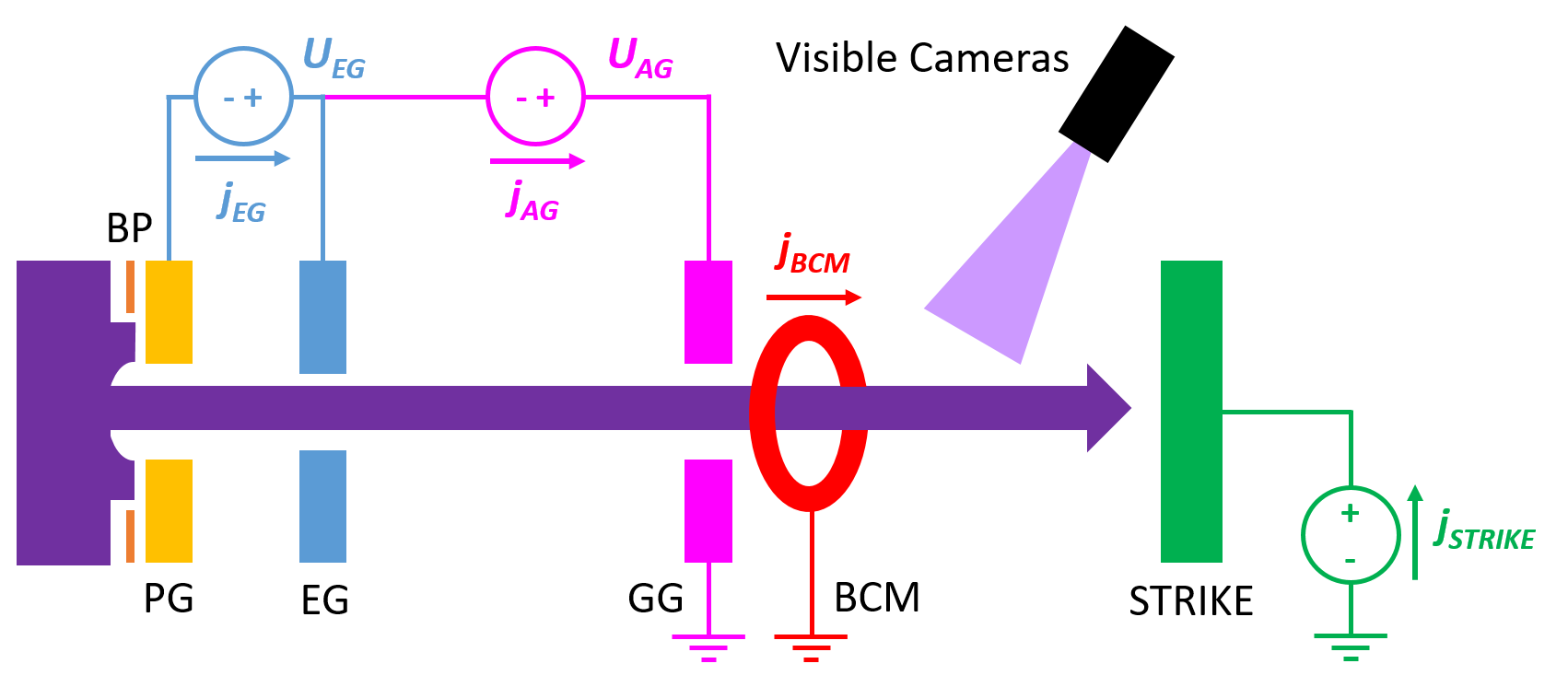}
\caption{Schematic of SPIDER accelerator showing the extraction and acceleration power supplies, and the positions of the Beamlet Current Monitors, visible cameras and STRIKE downstream of the grounded grid. Not to scale.}
\label{fig:Beam_diagnostic_schematic}
\end{figure}

In SPIDER negative ions are extracted from the source by the extraction potential U$_{EG}$ applied between the Plasma Grid (PG), which encloses the source, and the Extraction Grid (EG). The negative ions are then accelerated by the acceleration potential U$_{AG}$ applied between the EG and the Grounded Grid (GG). U$_{EG}$ + U$_{AG}$ are applied such that the source is at negative voltage, up to -108 kV (U$_{EG} = $-12 kV + U$_{AG}$ = -96kV). Electrons are co-extracted from the source and are deflected onto the EG by embedded co-extracted electron suppression magnets (CESMs). To reduce the electron temperature in the extraction region a transverse filter field is generated by a vertical current through the PG I$_{PG}$. The PG and a dedicated bias plate (BP) are biased positively with respect to the source to reduce the co-extracted electron current. The extraction and acceleration power supplies, named ISEG PS and AGPS, and the potentials applied to the three electrostatic grids are shown in figure~\ref{fig:Beam_diagnostic_schematic}. The ISEG and AGPS drain currents j$_{EG}$ and j$_{AG}$ provide estimations of the extracted ion current j$_{ex}$ (j$_{AG}$) and the co-extracted electron current (j$_{EG}$-j$_{AG}$).

The beam optics is determined by the beam perveance, defined by \eqref{eq:Perv}, and the electrostatic lens created by the extraction and acceleration potentials, with the strength of the lens given by U$_{AG}$/U$_{EG}$. 

\begin{equation}
\label{eq:Perv}
P = j_{ex}/U_{EG}^{3/2}
\end{equation}

At perveance match and with an optimum ratio U$_{AG}$/U$_{EG}$ the beam should be transmitted through the accelerator with a low divergence. Any large deviations in perveance or U$_{AG}$/U$_{EG}$ can result in an increasing beam divergence and direct interception on the grids. 

\begin{figure}[htpb]
\centering
\includegraphics[width=1\textwidth]{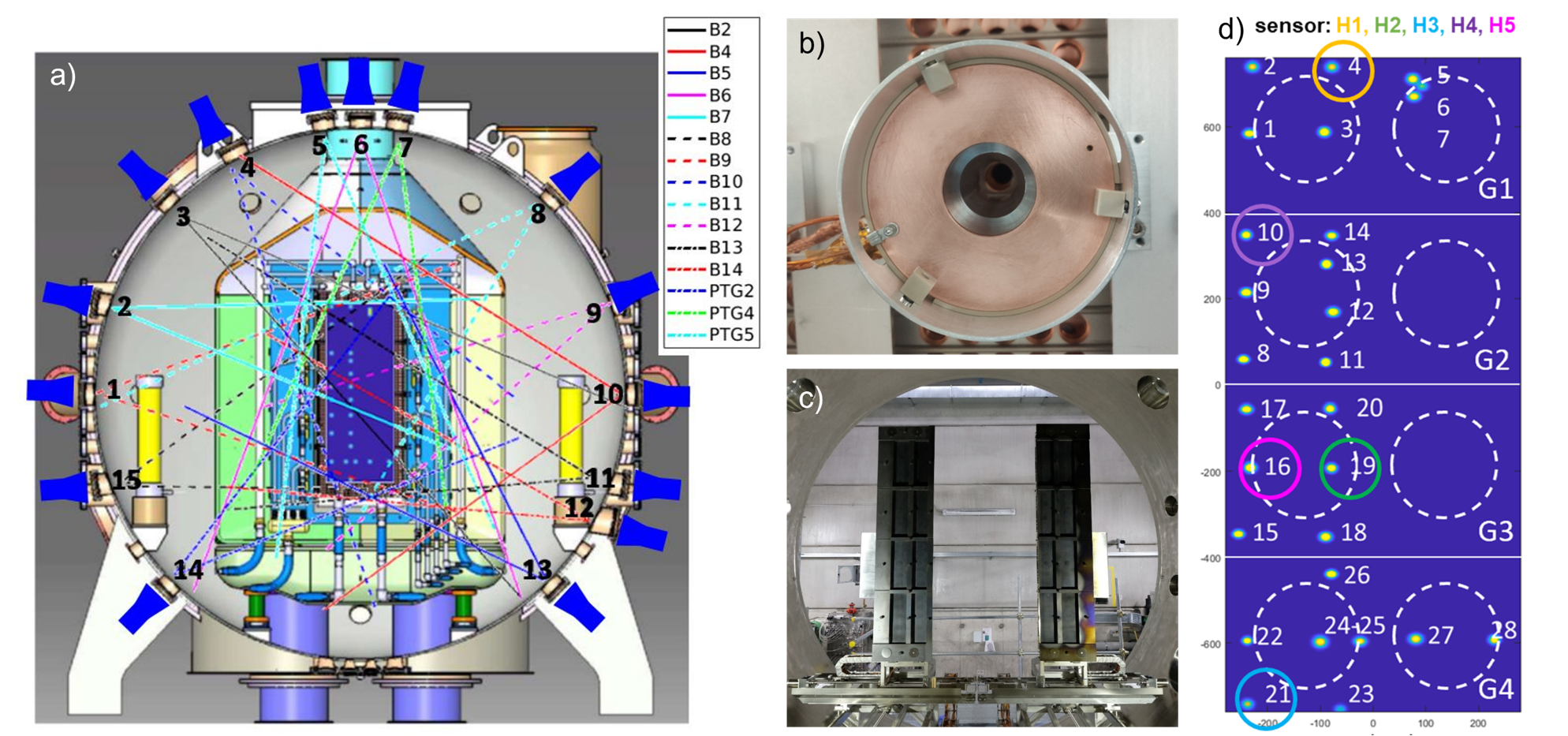}
\caption{Three SPIDER diagnostics that measure individual beamlets. a) Visible camera Lines of Sight used for tomography, b) single BCM mounted downstream of grounded grid and c) STRIKE calorimeter with 16 CFC tiles. d) 28 beamlets available during recent SPIDER experimental campaign. The BCM measures the five circled beamlets (H1-H5).}
\label{fig:Beam_diagnostics}
\end{figure}

With the PG mask reducing the number of open apertures it is possible to resolve the current and optics of individual beamlets. SPIDER has three diagnostics that can measure the same beamlets; BCMs, visible camera tomography and the STRIKE calorimeter (the relative positions of the diagnostics are given in figure~\ref{fig:Beam_diagnostic_schematic}). The BCMs are five sensors mounted downstream of the GG (shown in figure~\ref{fig:Beam_diagnostics}b) that measure the accelerated current of five individual beamlets (coloured circles in figure~\ref{fig:Beam_diagnostics}d). Downstream of the GG 15 cameras are positioned around the vacuum vessel to collect the visible light emitted from the beam interaction with the background gas (camera Lines of Sight shown in figure~\ref{fig:Beam_diagnostics}a). Using tomographic reconstruction the intensity (proportional to the current) and width of all the individual beamlets can be obtained. The STRIKE calorimeter, separated into two halves vertically, can be positioned to intercept the whole SPIDER beam or either the left or right halves (figure~\ref{fig:Beam_diagnostics}c), with the non-intercepted beam continuing on to the SPIDER beam dump. For the recent campaigns the left half of STRIKE (looking downstream from the source) intercepted 23 of the 28 available beamlets. The beamlet footprints are resolved by IR calorimetry on the CFC tiles; from this the beamlet current, width and position is determined. In addition, each tile is biased positively to collect the secondary electrons emitted due to the beam impact on the tiles. This provides an electrical measurement of the negative ion current intercepted by STRIKE (j$_{STRIKE}$ in figure~\ref{fig:Beam_diagnostic_schematic}). Of the five remaining beamlets three (5,6,7 in figure~\ref{fig:Beam_diagnostics}d) are measured by the AES, but the AES is not included in this comparison as they are different beamlets to the BCM and STRIKE. BES has also been excluded from this comparison. 

\section{Beam current comparison}

During 2021 cesium was evaporated into the SPIDER source for the first time, from three Cs ovens \cite{RIZZOLO20131007} placed vertically between horizontal driver pairs on the source back wall. The campaign was carried out in multiple phases; an exploration of the caesiation parameters at low (phase 1) and higher RF power (phase 2), beam optics studies (phase 3), HV testing at higher power (phase 4) and deuterium operation (phase 5) \cite{Sartori_2022}.

\begin{figure}[htpb]
\centering
\includegraphics[width=0.6\textwidth]{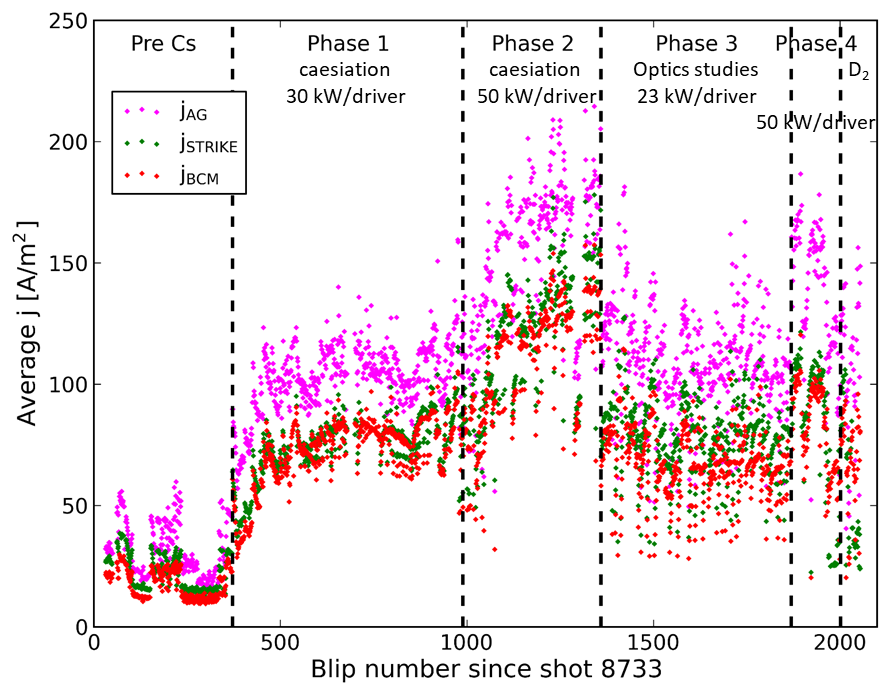}
\caption{SPIDER Cs campaign AGPS drain current (pink), BCM current (red) and STRIKE electrical current (green), averaged for a single beamlet.}
\label{fig:S21_beam_current}
\end{figure}

With the change from volume dominant to surface dominant H$^-$ production the extracted ion current increased five-fold and the electron-to-ion ratio decreased to <1 within 110 plasma pulses. The average beam current measured by the AGPS, BCM and STRIKE show a clear increase with the introduction of Cs (figure~\ref{fig:S21_beam_current}), as well as a significant difference between j$_{AG}$ and the accelerated current measured by the BCM and STRIKE. At 50 kW/driver j$_{AG}$ reaches 180 A/m$^{2}$, with the BCM and STRIKE lower at around 140-150 A/m$^{2}$. Due to a decrease in vacuum quality in phase 4 the currents do not return to their highest values at 50 kW/driver.

\begin{figure}[htpb]
\centering
\includegraphics[width=0.7\textwidth]{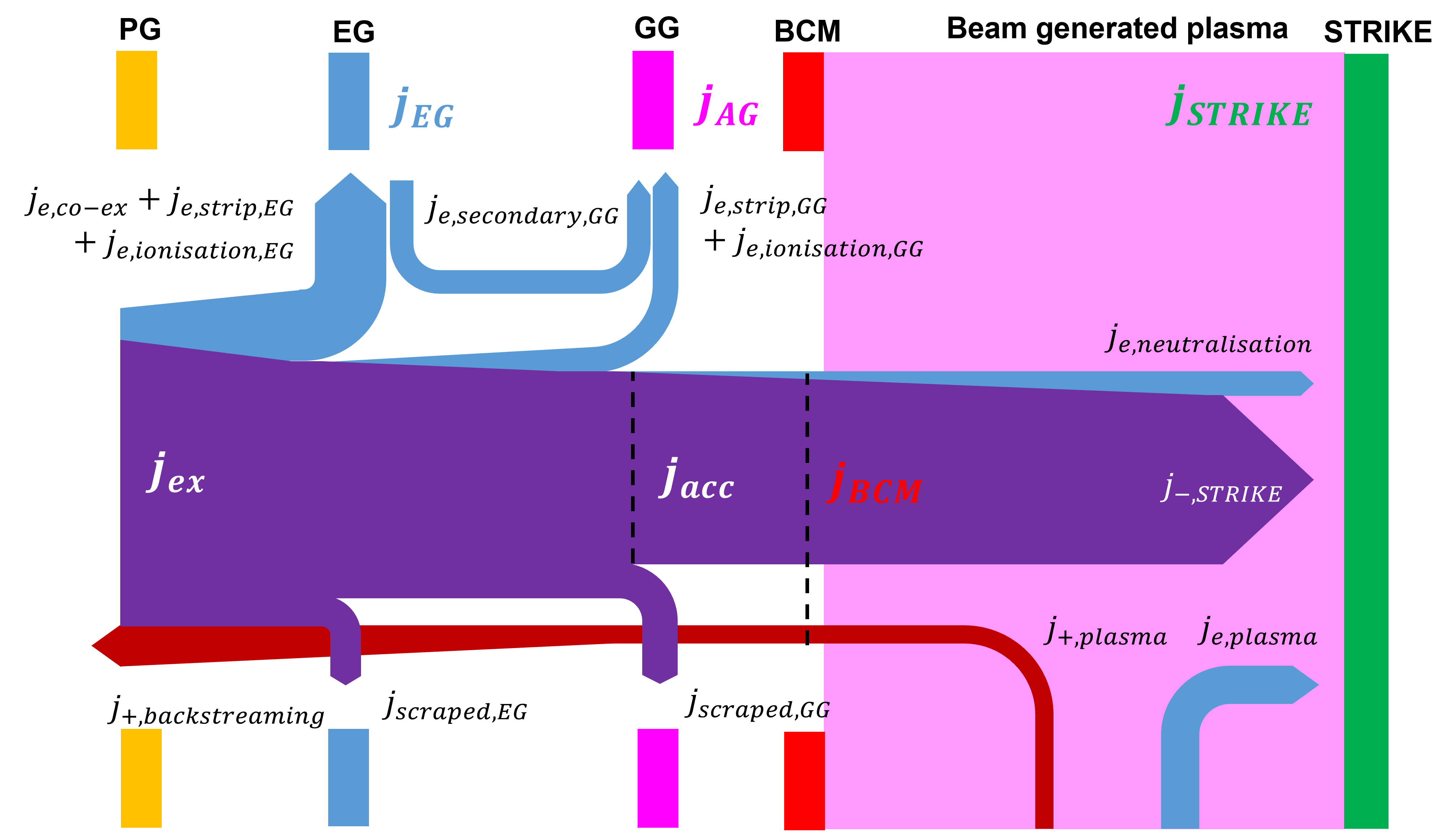}
\caption{Schematic of beam losses through the SPIDER accelerator, with the measurement points of the AGPS, BCM and STRIKE. Not to scale.}
\label{fig:beam_losses}
\end{figure}

The differences between the three measurements can be explained by the relative locations of the diagnostics and the secondary charges produced in the accelerator due to the background gas (volume processes) or beam scraping (surface processes) that contribute to the measurements (figure~\ref{fig:beam_losses}). At the PG the beam current is composed of extracted H$^{-}$ j$_{ex}$ and co-extracted electrons j$_{e,co-ex}$. Stripping and scraping of the beam results in an accelerated current j$_{acc}$ at the GG given by \eqref{eq:j_acc}.

\begin{equation}
\label{eq:j_acc}
j_{acc} = j_{ex}-j_{e,strip}-j_{scraped}
\end{equation}

The stripping, scraping and ionisation, with the distribution of charges onto the different grids, result in j$_{EG}$ and j$_{AG}$ being subtly different from j$_{ex}$, j$_{e,co-ex}$ and j$_{acc}$. j$_{EG}$ \eqref{eq:j_EG} is the sum of all the contributions in the accelerator, including electrons and backstreaming positive ions due to ionisation of the background gas. j$_{AG}$ \eqref{eq:j_AG} is the accelerated current plus some contribution from the stripping, ionisation and scraping (through direct interception on the GG and secondary electrons from the EG) in the accelerator, denoted here by the GG subscript.

\begin{equation}
\label{eq:j_EG}
j_{EG} = j_{ex}+j_{e,co-ex}+j_{e,ionisation}+j_{+,backstreaming}
\end{equation}
\begin{equation}
\label{eq:j_AG}
j_{AG} = j_{acc}+j_{e,strip,GG}+j_{e,secondary,GG}+j_{e,ionisation,GG}+j_{scraped,GG}
\end{equation}

The BCM and STRIKE both measure closer to j$_{acc}$. The BCM measures j$_{acc}$ almost directly \eqref{eq:j_BCM}, with a small additional contribution from backstreaming beam plasma positive ions. STRIKE collects the incident H$^{-}$ and electrons born from in-vessel neutralisation (j$_{e,neut}$, which are assumed to go in the beam direction and intercept STRIKE's large surface area; these two contributions should equal j$_{acc}$. Positively biased to recollect electrons emitted from its surface due to H$^{-}$ impact, STRIKE also collects the electrons from the beam generated plasma, which results in an over estimation of the accelerated current \eqref{eq:j_STRIKE}. This is seen in the slightly higher average current measured by STRIKE compared to the BCM. Both diagnostics are significantly lower than j$_{AG}$. The following subsections attempt to characterise these differences.

\begin{equation}
\label{eq:j_BCM}
j_{BCM} = j_{acc}+j_{+,plasma}
\end{equation}
\begin{equation}
\label{eq:j_STRIKE}
j_{STRIKE} = j_{-,STRIKE}+j_{e,neutralisation}+j_{e,plasma}
\end{equation}

\subsection{Volume processes}

\begin{figure}[htpb]
\centering
\includegraphics[width=1\textwidth]{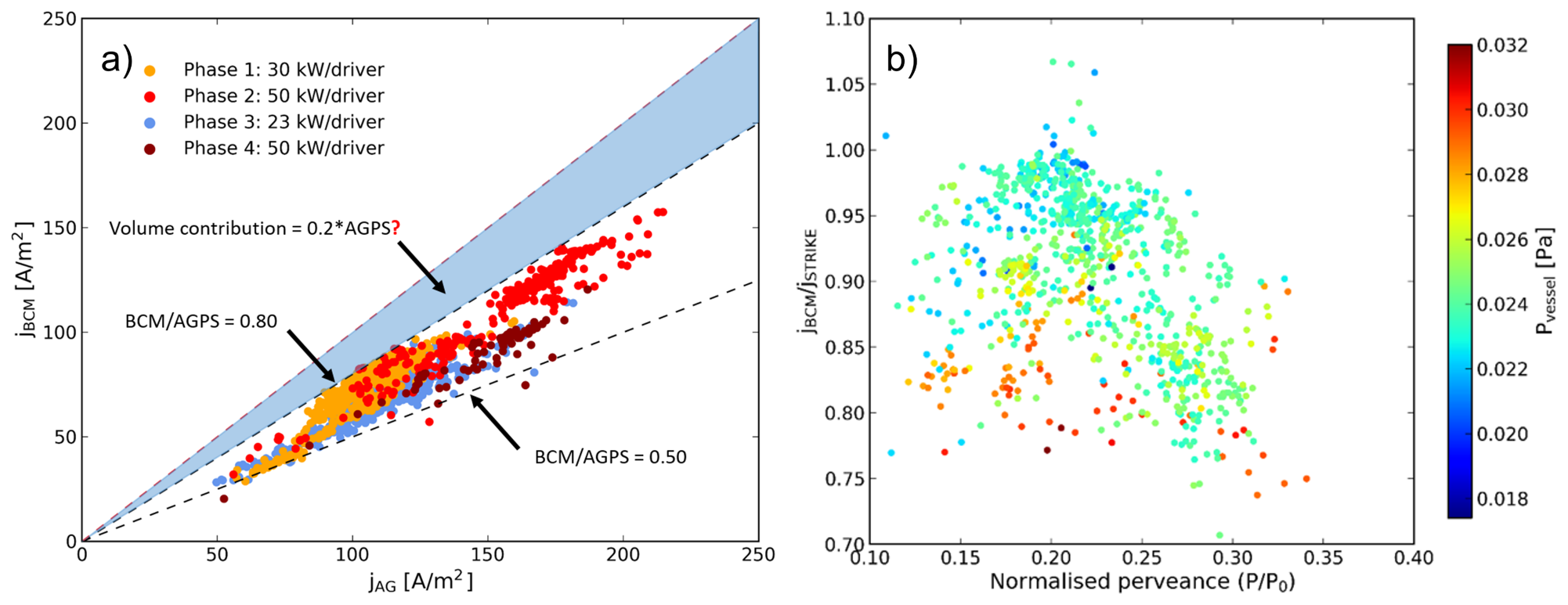}
\caption{a) Average BCM current against AGPS current in the four main phases of the Cs campaign. b) BCM/STRIKE against normalised beamlet perveance with SPIDER vessel pressure as the colourbar.}
\label{fig:BCM_comparison}
\end{figure}

First comparing the BCM and AGPS measurements (figure~\ref{fig:BCM_comparison}a) there is a minimum 20\% difference between the two. This is attributed to the volume contributions to j$_{AG}$, namely the stripping and ionisation in the accelerator. No clear dependence of the current ratio on the source or vessel pressure is seen; most of the campaign was performed with a source filling pressure p$_{fill}$ around 0.35 Pa. Estimations of the stripping losses with the PG mask are around 7\% of the extracted current \cite{SARTORI2023113730}, with BES analysis giving similar numbers \cite{Agnello_2022}. Considering that ionisation will also contribute to the current, although to a smaller degree, 20\% is a high but not unrealistic estimation of the volume processes, although it would require all of the additional charges to be collected by the AGPS and not shared with the ISEG PS. The remaining difference in BCM/AGPS we will characterise in the subsequent sections. 

The ratio STRIKE/AGPS (not shown) is 5-10\% higher than BCM/AGPS, attributed to the electrons collected from the beam plasma in the source. The BCM/STRIKE ratio shows a far smaller variation, as both are measuring currents close to j$_{acc}$. There is some dependence on vessel pressure as expected (figure~\ref{fig:BCM_comparison}b); the main difference is the beam plasma, which is generated from the background gas in the vessel. The perveance is calculated using j$_{AG}$, as it is the closest to j$_{ex}$. From OPERA simulations optimum perveance is around 0.26P$_{0}$ for j$_{ex}$ = 355 A/m$^{2}$ and U$_{EG}$ = 9.4 kV \cite{Agostinetti_2011}. P$_{0}$ is the perveance limit for a planar diode, given by \eqref{eq:P0} where d$_{PGEG}$ is 6 mm.

\begin{equation}
\label{eq:P0}
P_0 = 4/9\epsilon_0\sqrt{2e/m_H}(1/d_{PGEG})^2
\end{equation}

\subsection{Initial ceasiation}

\begin{figure}[htpb]
\centering
\includegraphics[width=1\textwidth]{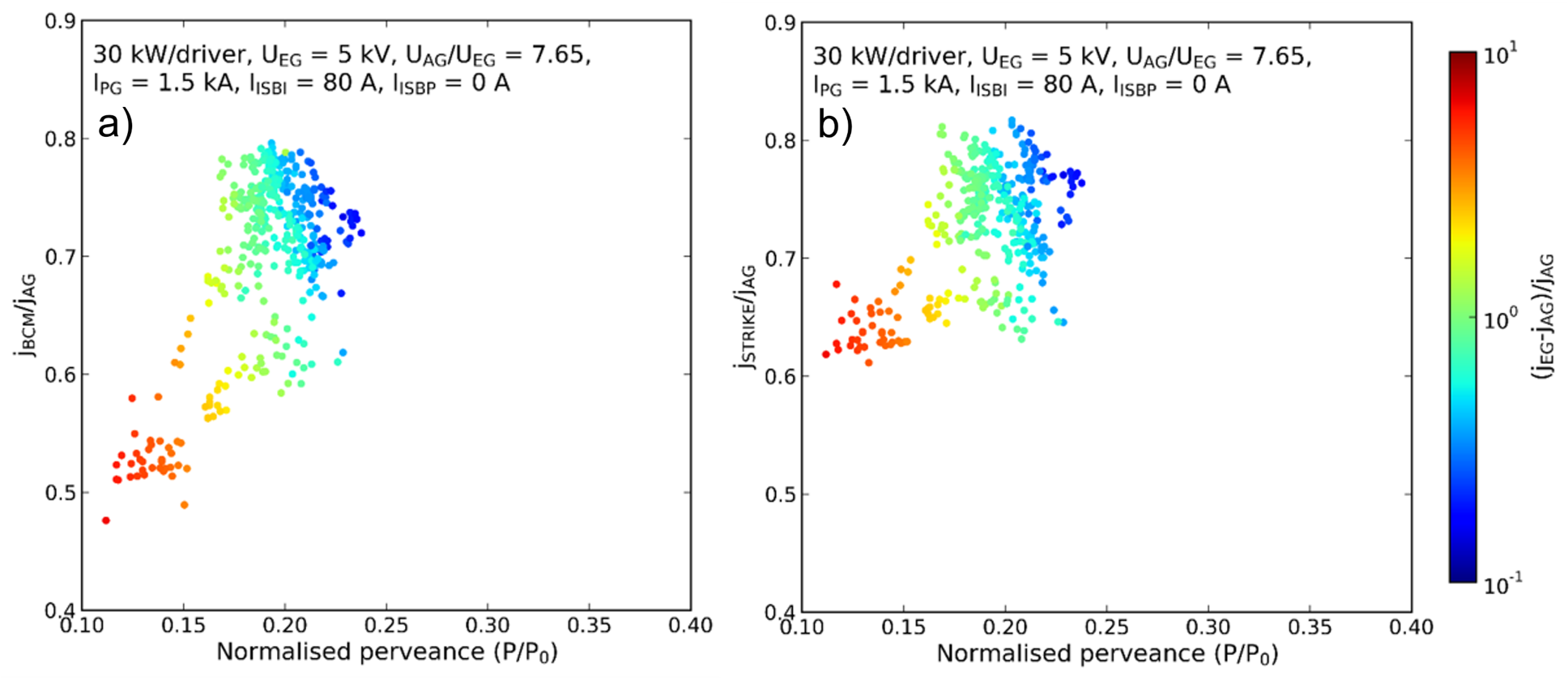}
\caption{a) BCM/AGPS and b) STRIKE/AGPS against normalised beamlet perveance for phase 1 pulses with fixed source parameters (orange points in figure~\ref{fig:BCM_comparison}a). The colorbar is an approximation of the electron to ion ratio.}
\label{fig:Phase1_perv}
\end{figure}

Taking the pulses from the initial ceasiation at 30 kW/driver (phase 1, orange points in figure~\ref{fig:BCM_comparison}a) the current ratio BCM/AGPS and STRIKE/AGPS both increase as the perveance increases towards 0.26P$_{0}$ (figure~\ref{fig:Phase1_perv}). The co-extracted electron to ion ratio decreases (using the difference between j$_{EG}$ and j$_{AG}$ for the co-extracted electrons) as the ratio improves. This improvement can be taken as an increase in the beam transmission as the optics improve with ceasiation. 

\subsection{Overperveance}

\begin{figure}[htpb]
\centering
\includegraphics[width=1\textwidth]{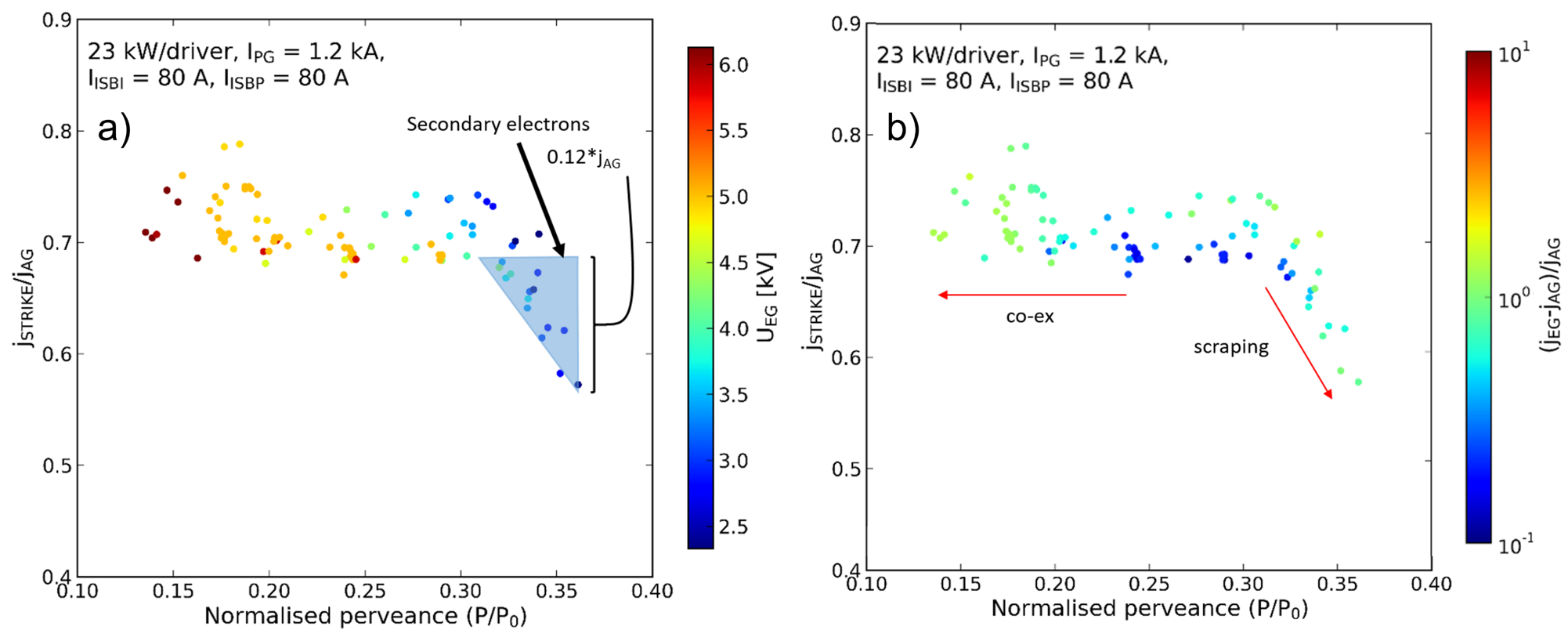}
\caption{STRIKE/AGPS against normalised beamlet perveance for phase 3 pulses with fixed source parameters (subset of the light blue points in figure~\ref{fig:BCM_comparison}a). The colorbar is a) the extraction voltage, b) an approximation of the electron to ion ratio.}
\label{fig:STRIKE_vs_AGPS_overperv}
\end{figure}

On the other hand, overperveance, where U$_{EG}$ is too low for the available current, results in a decrease in the current ratio. This can be seen in pulses taken from phase 3, where dedicated studies of the beam source parameters and optics were performed once the source was well caesiated. In figure~\ref{fig:STRIKE_vs_AGPS_overperv}, during extraction voltage scans the ratio of STRIKE/AGPS current drops considerably past 0.3P$_{0}$. This is due to scraping of the over-perveance beam onto the EG, resulting in secondary electrons that are collected by the AGPS but not by STRIKE (or the BCM). These secondary electrons make up to approximately 12\% of j$_{AG}$ at >0.35P$_{0}$. The difference between j$_{EG}$ and j$_{AG}$ increases as the perveance decreases, due to an increase in the co-extracted electron current. Over perveance, the difference also increases due to scraping on the EG.

\section{Individual beamlets}

\begin{figure}[htpb]
\centering
\includegraphics[width=01\textwidth]{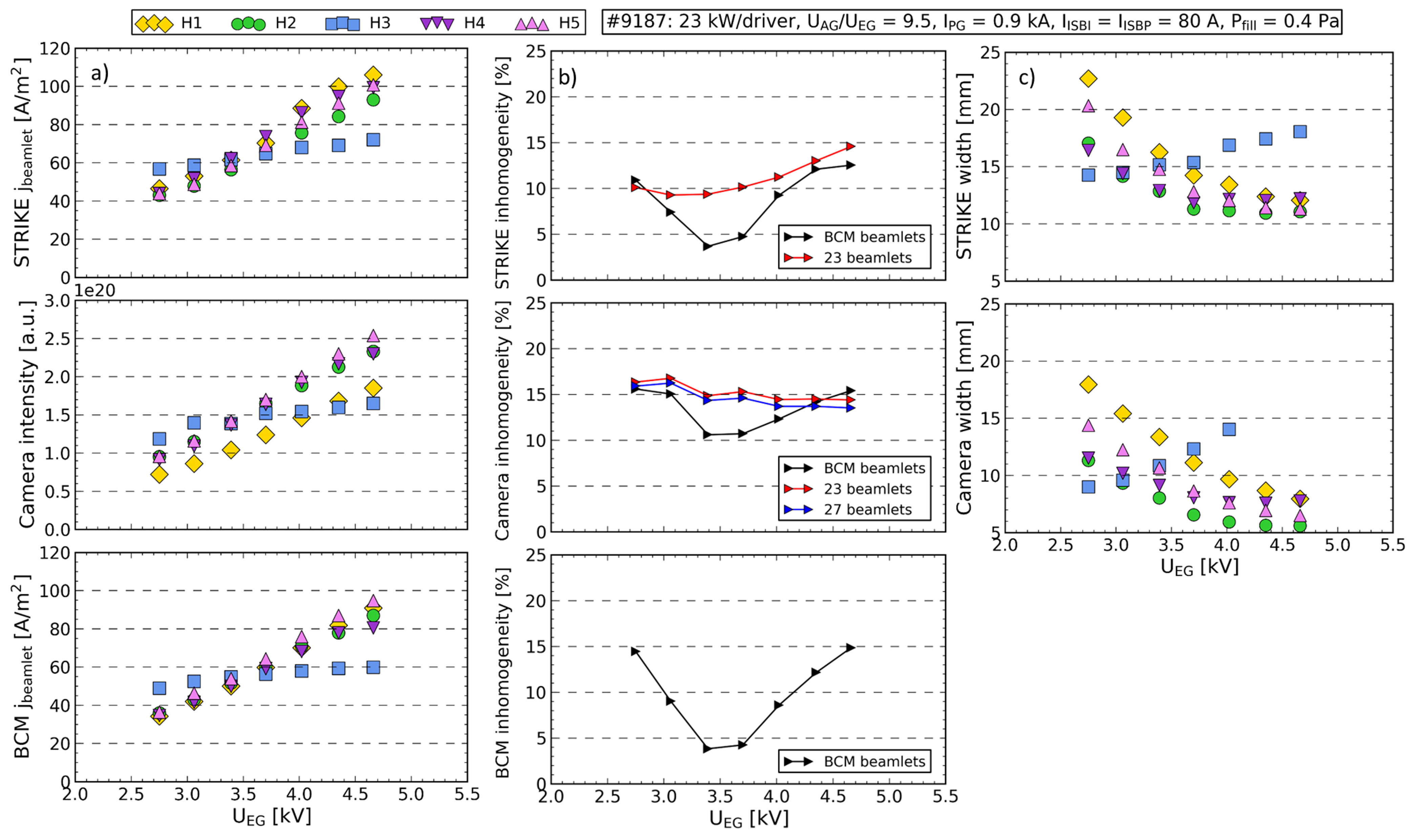}
\caption{Comparison between STRIKE calorimetry, tomography and BCM for pulse 9187. a)  Beamlet current (intensity for tomography) for the five BCM beamlets. b) Current inhomogeneity for the five BCM beamlets (black), all STRIKE beamlets (red) and all but one camera beamlets (blue). c) Beamlet width for the five BCM beamlets for STRIKE and tomography.}
\label{fig:Beamlets_comparison}
\end{figure}

The individual beamlets are compared using the BCM, STRIKE IR calorimetry and tomography, with the latter two providing information on the beamlet divergence as well as the current. The beamlet current for the five BCM beamlets follow the same trend with extraction voltage and for all three diagnostics (figure~\ref{fig:Beamlets_comparison}a). The current of each beamlet increases with U$_{EG}$, eventually reaching a saturation point where all the available H$^-$ is extracted. The saturation voltage is not reached in this case, but Shepherd et al. \cite{SHEPHERD2023113599} has shown that this occurs at U$_{EG}$ > 6 kV for this RF power. The inhomogeneity $\gamma$ in the beamlet current can be estimated as the standard deviation of the beamlet currents divided by the mean of the beamlet currents \eqref{eq:inhomogeneity}. 

\label{eq:inhomogeneity}
\begin{equation}
\gamma = \frac{\sigma_{j_{beamlets}}}{\mu_{j_{beamlets}}} \,
%\sigma = \frac{\sqrt{\frac{1}{N}\sum\limits_{i=1}^N (j_i - \mu)}}{\mu} \,
%\mu = \frac{1}{N}\sum\limits_{i=1}^N j_i
\end{equation}

For each diagnostic the current inhomogeneity of the five BCM beamlets has a minimum at the same U$_{EG}$ (figure~\ref{fig:Beamlets_comparison}b). It is worth noting that the increase in inhomogeneity either side of the optimum U$_{EG}$ is due to the different behaviour of the bottom beamlet H3. The bottom of the SPIDER source has a lower plasma density due to vertical drifts induced by the filter field, and a lower H$^-$ availability. A study of the beam plasma uniformity, and the resulting current uniformity, has been carried out by Serianni et al \cite{Serianni_2022}.

One difference between the three diagnostics is the intensity of the top beamlet H1, as measured by tomography, compared to the current measured by STRIKE and the BCM. For the tomography H1 is consistently lower than the other beamlets, resulting in a higher overall inhomogeneity. This is even more apparent when considering all of the available beamlets for the diagnostics (red and blue lines in figure~\ref{fig:Beamlets_comparison}b). For STRIKE with more beamlets there is less variation in the inhomogeneity with  U$_{EG}$, but the overall trend does not change. For the cameras the inhomogeneity is higher and no longer has a trough, due to a larger variation between beamlet intensities at all U$_{EG}$ compared to STRIKE. The camera calibrations may explain the wider range of intensities, in particular the lower intensity of beamlet H1. For the tomography the central beamlet of the vertical group of three has been excluded (beamlet 6 in figure~\ref{fig:Beam_diagnostics}) as it has a much lower intensity than the other three. This may be due to the difficulty in isolating the beamlet from it's neighbours for the tomographic reconstruction. Its inclusion results in an increase in the blue inhomogeneity line of over 5\%.

The beamlet width, calculated from STRIKE and tomography, has a minimum where the current deviates from the Child Langmuir law (figure~\ref{fig:Beamlets_comparison}c), i.e. where the current saturates due to H$^-$ availability (generally >5 kV for I$_{PG}$ = 0.9 kA). The only exception is H3, which for both diagnostics has a minimum width at much lower U$_{EG}$. Combined with the higher current at low U$_{EG}$ the implication is that the other beamlets may be scraping on the EG, which reduces the measured accelerated current. 

\begin{figure}[htpb]
\centering
\includegraphics[width=1\textwidth]{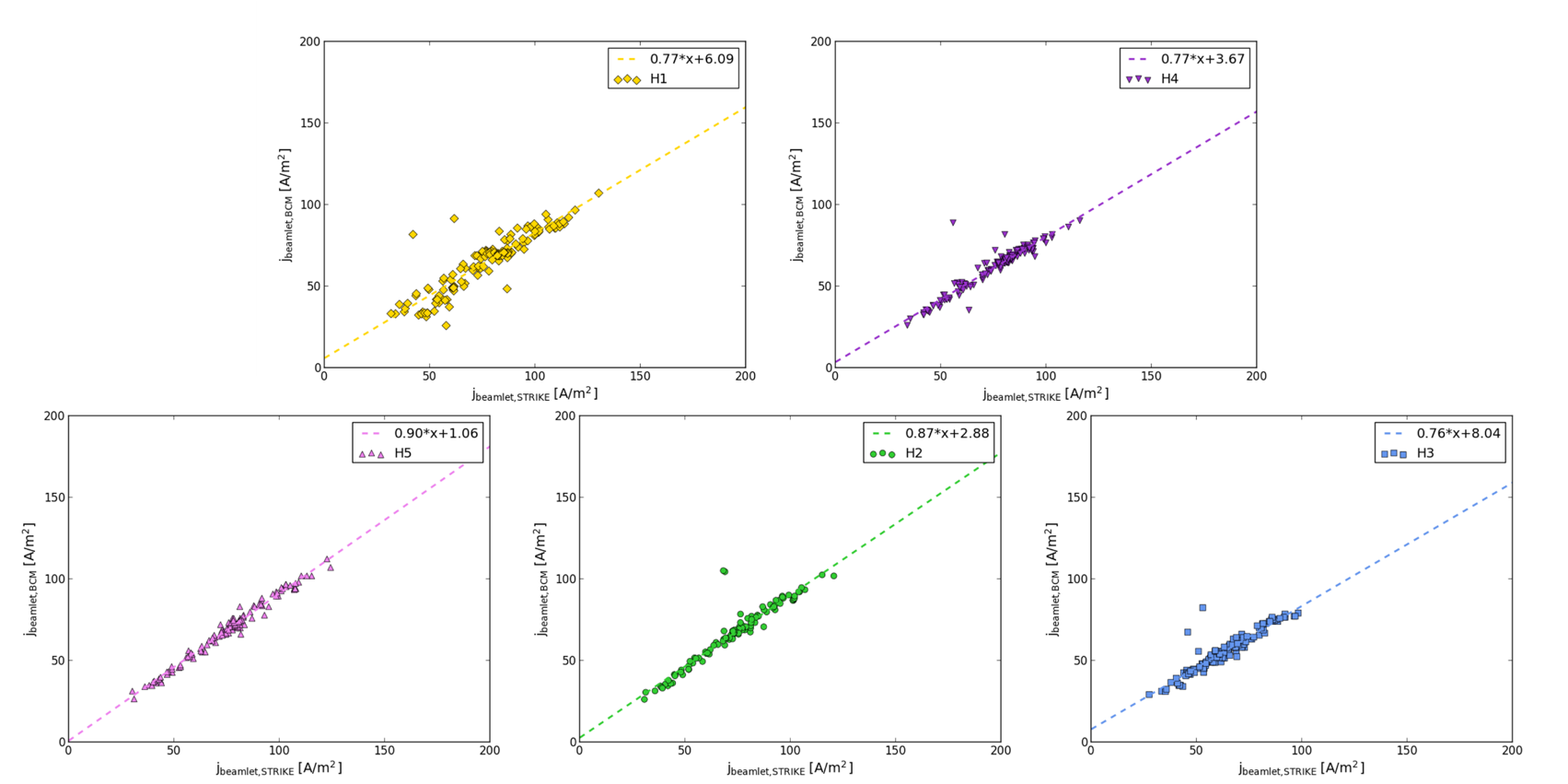}
\caption{BCM current against STRIKE IR current for the five BCM beamlets. Points taken from phase 3 optics scans at low RF power (light blue points in figure~\ref{fig:BCM_comparison}a).}
\label{fig:BCM_vs_STRIKE_beamlets}
\end{figure}

Taking all the pulses from the 23 kW/driver optics studies the individual beamlets compare consistently between the BCM and STRIKE IR current measurements (figure~\ref{fig:BCM_vs_STRIKE_beamlets}), with a BCM/STRIKE ratio ranging from 0.76 to 0.9. The variation between the beamlets is likely due to the varying emmissivity of the STRIKE CFC tiles, both from tile to tile and across each tile. However, within the source parameter variations in phase 3 the BCM and STRIKE IR current ratios are unchanged.

\section{Conclusion}

With the reduction of the number of open apertures on SPIDER it has been possible to measure the current of some of the apertures directly using the BCM. The current measurements compare well with existing diagnostics that can resolve the same individual beamlets. Comparing the average BCM current to the STRIKE electrical current, both provide a more accurate measurement of the accelerated current than the AGPS drain current. The latter is affected by secondary charges, with additional work needed to quantify these components. 

Plans to increase the number of beamlets measured by the BCM for the next SPIDER campaign with the PG mask will improve the accuracy of the current comparison. A study into the design of beamlet group current sensors, for when the PG mask is removed, is also underway. 

\acknowledgments

This work has been carried out within the framework of the ITER-RFX Neutral Beam Testing Facility (NBTF) Agreement and has received funding from the ITER Organization. The views and opinions expressed herein do not necessarily reflect those of the ITER Organization.

This work has been carried out within the framework of the EUROfusion Consortium, funded by the European Union via the Euratom Research and Training Programme (Grant Agreement No 101052200 — EUROfusion). Views and opinions expressed are however those of the author(s) only and do not necessarily reflect those of the European Union or the European Commission. Neither the European Union nor the European Commission can be held responsible for them.

This work was supported in part by the Swiss National Science Foundation.

\bibliographystyle{JHEP.bst}
\bibliography{mybib}

\end{document}